\documentclass{iau_FM}
\usepackage{graphicx,natbib}

\newcommand{\be}{\begin{equation}}
\newcommand{\ee}{\end{equation}}

\newcommand{\bea}{\begin{eqnarray}}
\newcommand{\eea}{\end{eqnarray}}
\newcommand{\bean}{\begin{eqnarray*}}
\newcommand{\eean}{\end{eqnarray*}}

\newcommand{\EQ}{\begin{equation}}
\newcommand{\EN}{\end{equation}}
\newcommand{\ba}{\begin{eqnarray}}
\newcommand{\ea}{\end{eqnarray}}

\graphicspath{{./fig/}{./png/}}

\title[Magnetism in the Early Universe]
{Magnetism in the Early Universe}

\author[Kahniashvili et al.]
{Tina Kahniashvili$^{1,2}$
Axel Brandenburg$^{3,4,1}$, Arthur Kosowsky$^{5}$,\\
Sayan Mandal$^{1}$, Alberto Roper Pol$^{4,6}$ }

\affiliation{$^1$Department of Physics, Carnegie Mellon University, USA
\\
$^2$Abastumani Astrophysical Observatory,
Ilia State University, Georgia
\\
$^3$Nordita, KTH Royal Institute of Technology \& Stockholm University;
Department of Astronomy, Stockholm University, Sweden;
and JILA, University of Colorado at Boulder, USA
\\
$^4$Laboratory for Atmospheric and Space Physics,
University of Colorado at Boulder, USA
\\
$^5$Department of Physics and Astronomy and PITT PACC, University of Pittsburgh, USA
\\
$^6$Department of Aerospace Engineering,
University of Colorado at Boulder, USA
}

\pubyear{2018}
\jname{Astronomy in Focus}
\editors{Teresa Lago, ed.}
\begin{document}

\maketitle

\begin{abstract}
Blazar observations point toward the possible presence of magnetic fields
over intergalactic scales of the order of up to $\sim1\,$Mpc, with
strengths of at least $\sim10^{-16}\,$G.
Understanding the origin of these large-scale magnetic fields is a
challenge for modern astrophysics.
Here we discuss the cosmological scenario, focussing on the following questions:
(i) How and when was this magnetic field generated?
(ii) How does it evolve during the expansion of the universe?
(iii) Are the amplitude and statistical properties of this field such that they can
explain the strengths and correlation lengths of observed magnetic fields?
We also discuss the possibility of observing primordial turbulence through direct
detection of stochastic gravitational waves in the mHz range accessible to LISA.


\end{abstract}

\firstsection 
\section{Introduction}

Magnetic fields of strengths of the order of $\sim 10^{-16}\,$G are thought to be
present in the voids between galaxy clusters; see
\cite{Neronov:1900zz} for the pioneering work and \cite{Durrer:2013pga}
for a review and references therein.
These are thought to be the result of seed magnetic field amplification,
with two scenarios of the origin currently under discussion;
see \cite{Subramanian:2015lua} for a review:
a bottom-up (astrophysical) scenario, where the seed is typically very
weak and magnetic field is transferred from local sources within galaxies to larger scales,
and a top-down (cosmological) scenario where a magnetic field is generated prior
to galaxy formation in the early universe on scales that are large at the
present epoch.
We discuss two different primordial magnetogenesis scenarios: inflationary and cosmological phase transitions.
We address cosmic magnetohydrodynamic (MHD) turbulence, in order to understand the magnetic field evolution.
Turbulent motions can also affect cosmological phase transitions.
We argue that even a small total energy density in turbulence
(less than 10\% of the total thermal energy density),
can have substantial effects because of strong nonlinearity of the relevant physical processes;
see also \cite{Vazza:2017qge}.

\section{Overview}

The evolution of a primordial magnetic field
is determined by various physical processes that result in amplification as well as damping.
Complexities arise in the problem due to the strong coupling between magnetic field
and plasma motions \citep{Kahniashvili:2010gp},
producing MHD turbulence,
which then undergoes free decay after the forcing
is switched off \citep{Brandenburg:1996fc,Dimopoulos:1996nq,Jedamzik:1996wp,Subramanian:1998fn};
see \cite{Kahniashvili:2015msa} for a recent overview.
The presence of initial kinetic and/or magnetic helicity strongly affects the development of turbulence.
In several models of phase transition magnetogenesis, parity (mirror symmetry) violation
leads to a non-zero chirality (helicity) of the field \citep{Cornwall:1997ms,
Giovannini:1997eg,Field:1998hi,Giovannini:1999wv,Vachaspati:2001nb}.
We also underline the importance of possible kinetic helicity: our recent simulations
have shown that through the decay of hydromagnetic turbulence with initial kinetic helicity,
a weak nonhelical magnetic field eventually becomes
fully helical \citep{Brandenburg:2017neh}.

The anisotropic stresses of the resulting turbulent magnetic and kinetic fields
are a source of gravitational waves, as already pointed out by
\cite{Deryagin:1986qq}.
The amplitude of the gravitational wave spectrum depends on the strength of the turbulence,
and its characteristic wavelength depends on the energy scale at which the
phase transition occurs \citep{Gogoberidze:2007an}.

\section{Results}

Understanding the mechanisms for generating primordial turbulence
is a major focus of our investigation.
Turbulence may be produced during cosmological
phase transitions
when the latent heat of the phase transition is partially converted to
kinetic energy of the plasma as the bubbles expand, collide, and
source plasma turbulence \citep{Christensson:2000sp}.
The two phase transitions of interest in the early universe are (i) the electroweak
phase transition occurring at a temperature of $T \sim 100\,$GeV, and (ii) the
QCD phase transition occurring at $T \sim 150\,$MeV.
Turbulence at the electroweak phase transition scale is more interesting for the gravitational wave detection prospects, since
the characteristic frequency of the resulting stochastic gravitational wave background, set
by the Hubble length at the time of the phase transition,
falls in the Laser Interferometer Space Antenna (LISA)
 frequency band; see \cite{Kamionkowski:1993fg},
and \cite{Kosowsky:2001xp} for pioneering studies, and
\cite{Caprini:2018mtu} for a recent review.

Since the electroweak phase transition is probably a smooth crossover in the
Standard Model of particle physics, it would not proceed through
bubble collisions and follow up turbulence.
Our knowledge of electroweak scale physics is incomplete; at least
two lines of reasoning point toward a first-order phase transition in the very early universe.
First, such a transition can provide the out-of-equilibrium environment
necessary for successful baryogenesis; see, e.g., \cite{Morrissey:2012db}.
Secondly, as discussed above, turbulence induced in a first-order transition
naturally amplifies seed magnetic fields which can explain
the magnetic fields that might be present in cosmic voids; see Fig.~1 and
\cite{Brandenburg:2017neh}.
Arguments in favor of a primordial origin of such fields
were also given by \cite{Dolag:2010ni}.

\begin{figure}[t!]
\begin{center}
\includegraphics[width=0.7\textwidth]{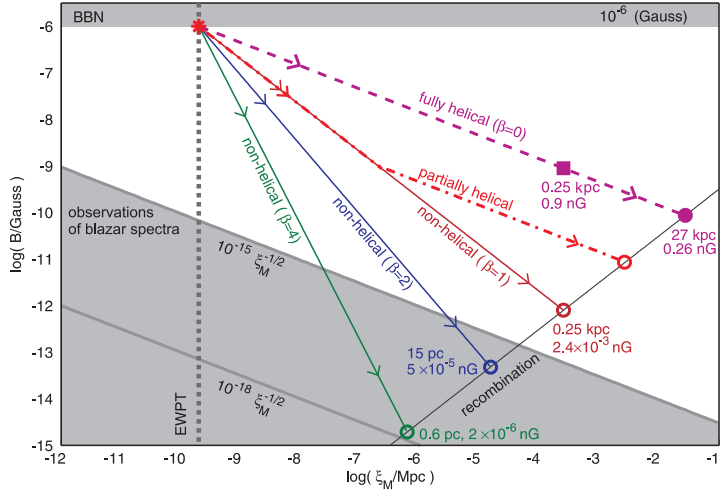}
\end{center}
\caption[]{
Turbulent evolution of the strength $B_{\rm rms}$ and correlation length $\xi_M$
of the magnetic field starting from their
upper limits given by the Big Bang Nucleosynthesis (BBN) 
bound and the horizon scale at the electroweak phase transitions
(from \cite{Brandenburg:2017neh}, Fig.~11).
}
\label{Cons2}
\end{figure}

\begin{figure*}[t!]\begin{center}
\includegraphics[width=.28\textwidth]{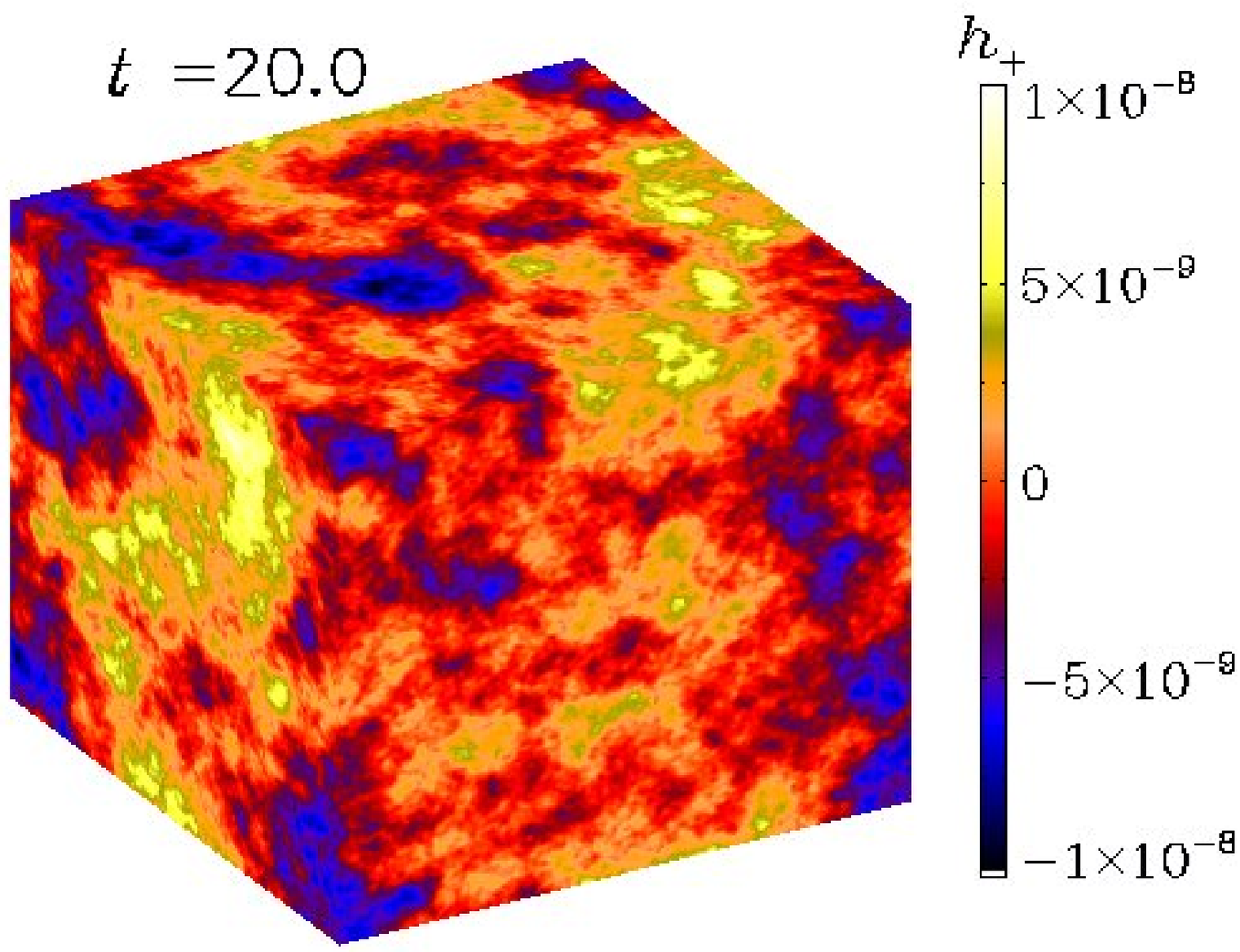}
\includegraphics[width=.28\textwidth]{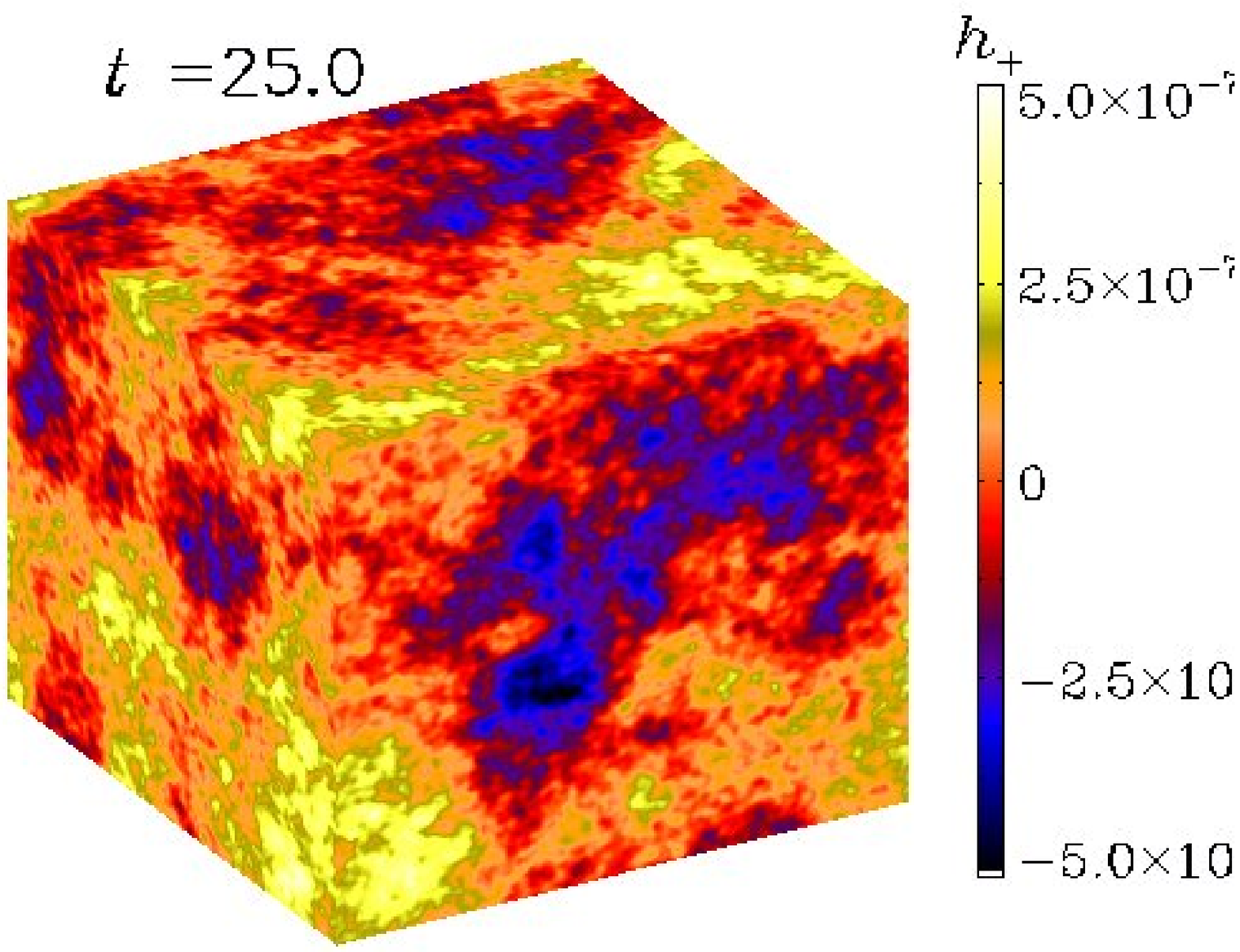}
\includegraphics[width=.28\textwidth]{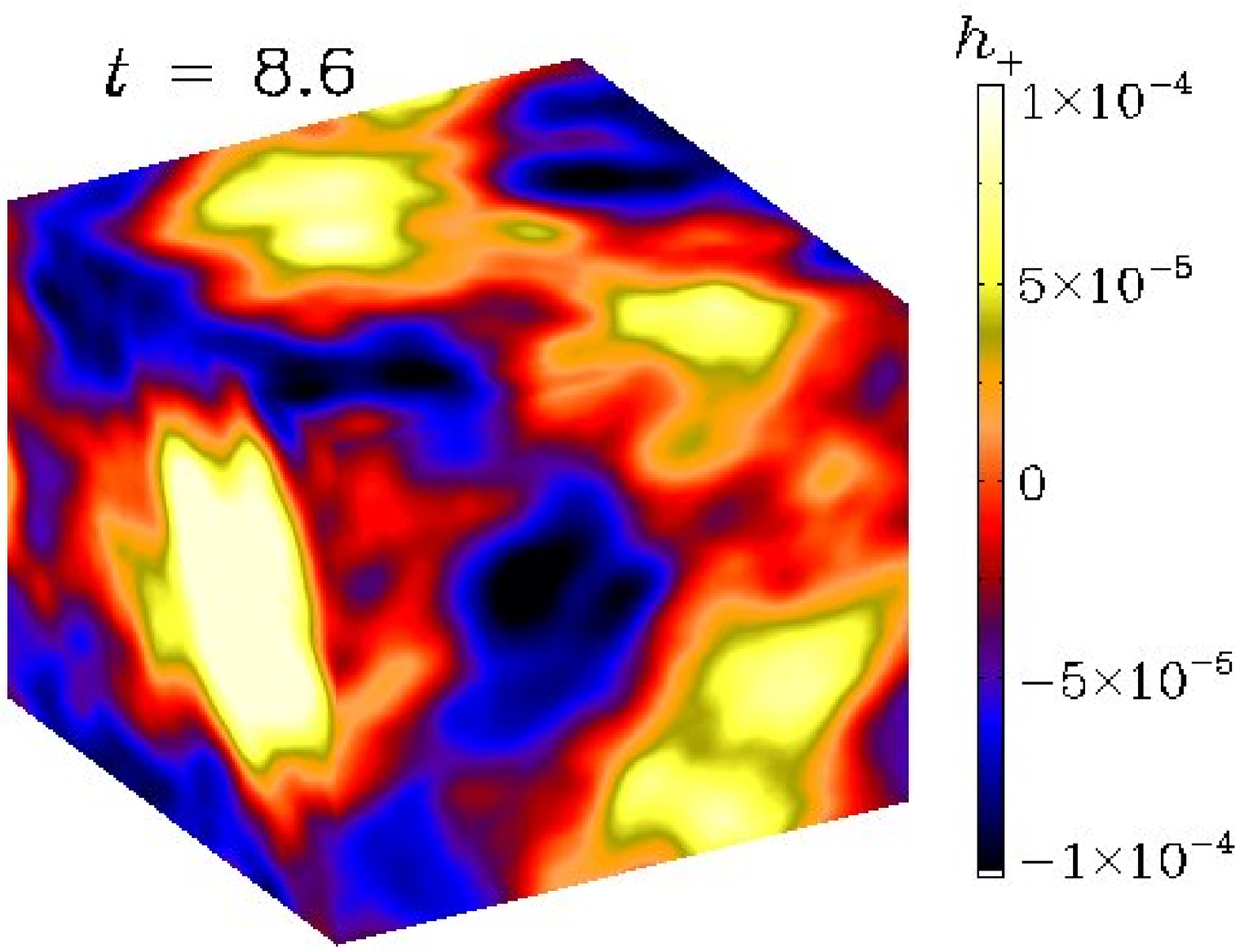}
\includegraphics[width=.28\textwidth]{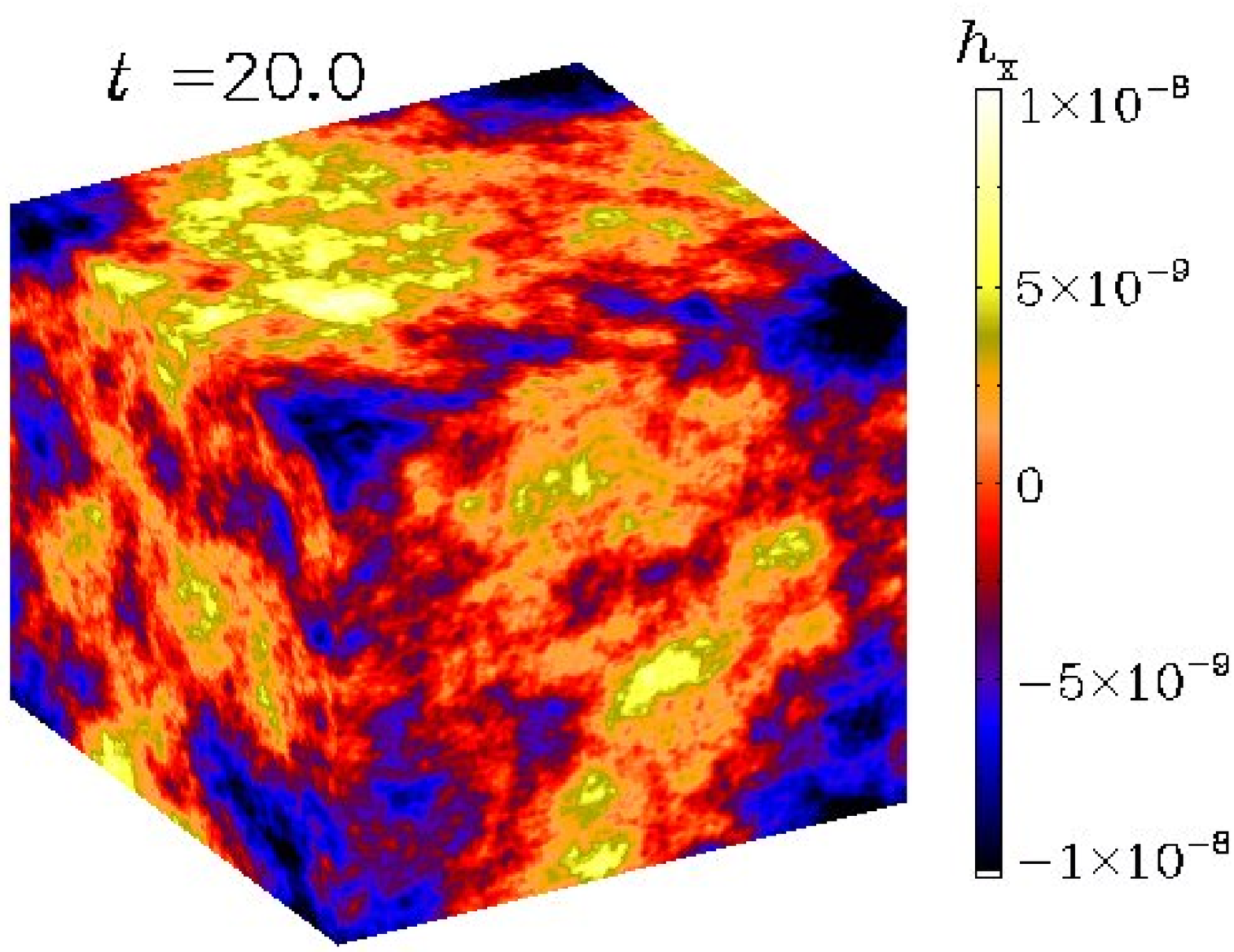}
\includegraphics[width=.28\textwidth]{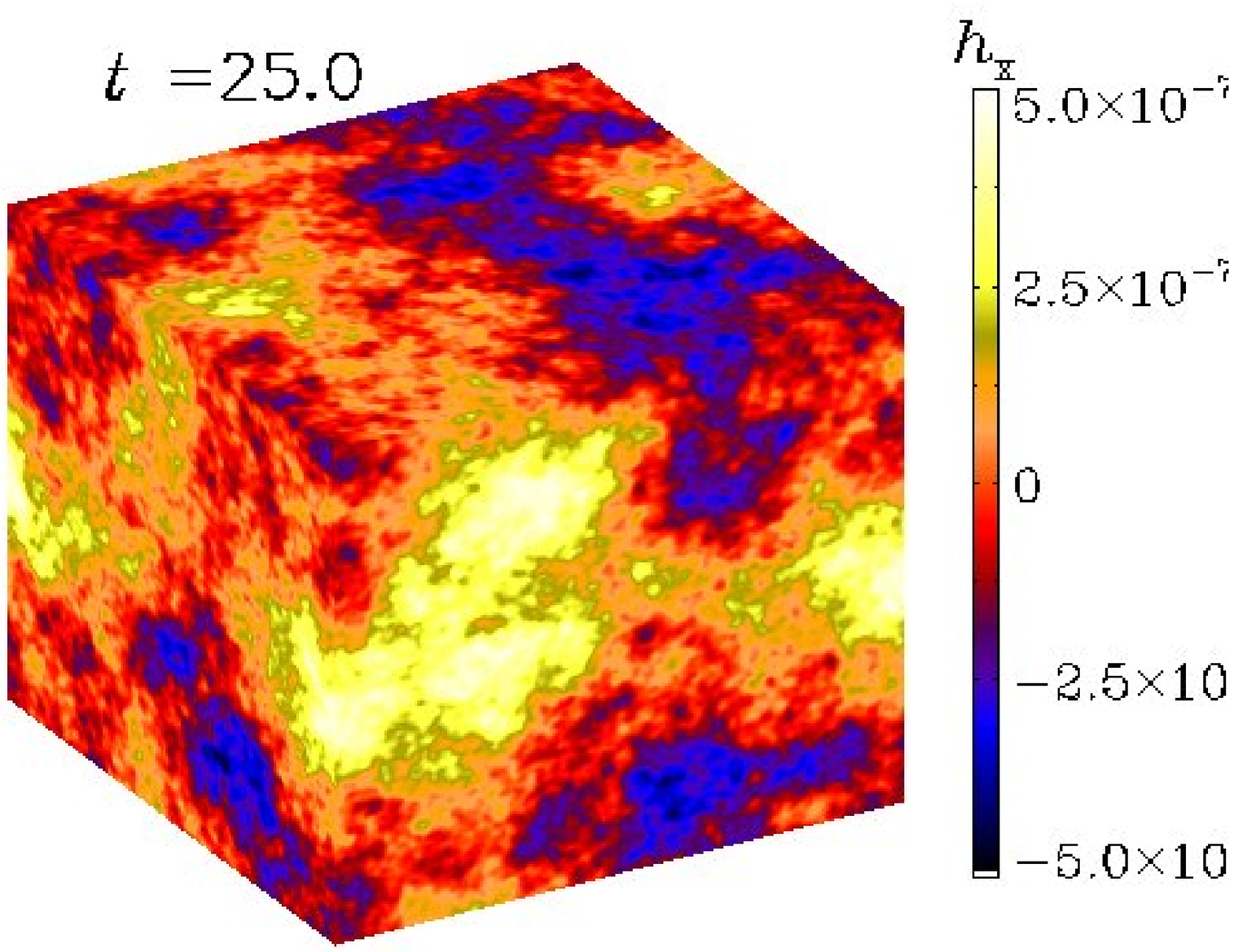}
\includegraphics[width=.28\textwidth]{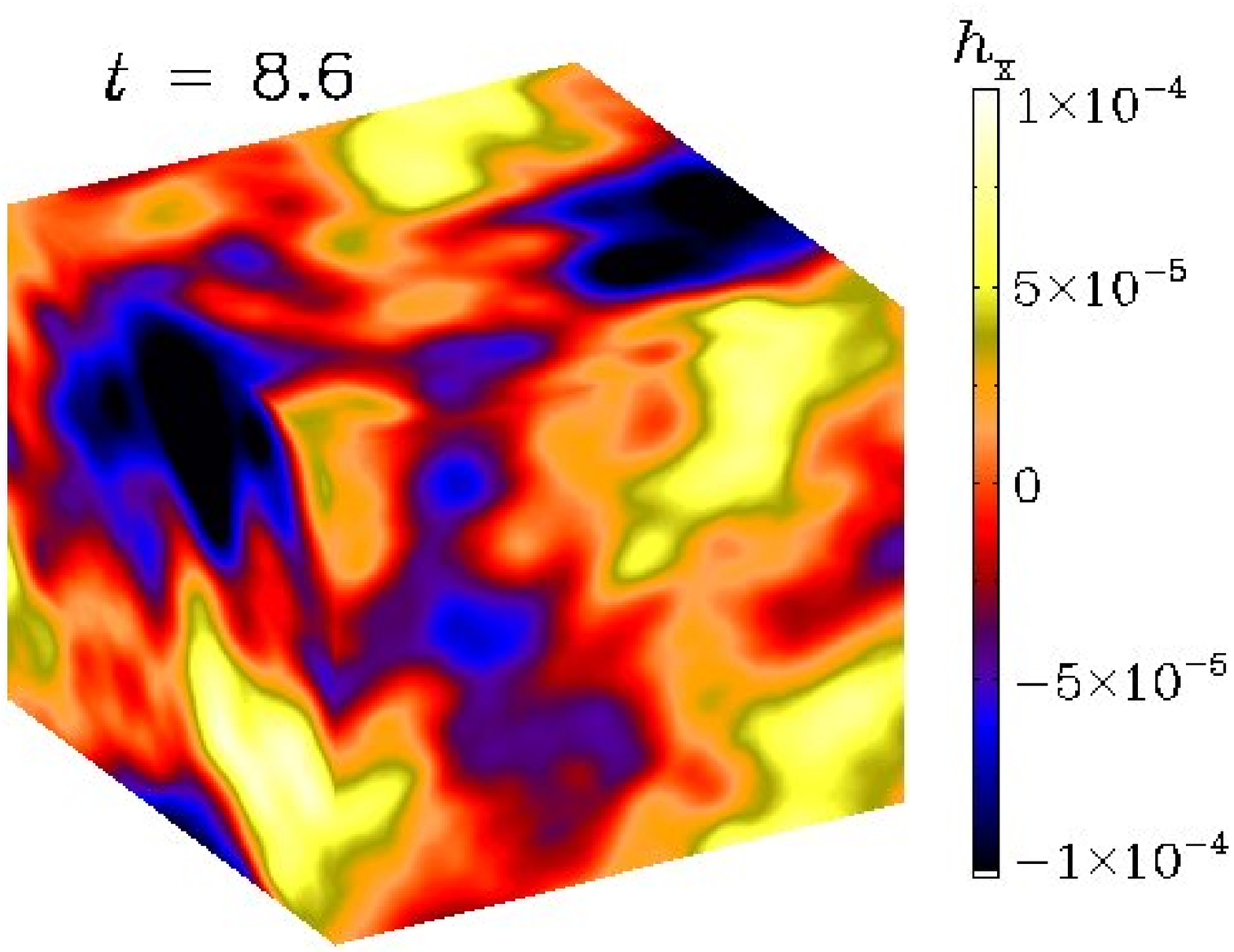}
\end{center}\caption[]{
Visualizations of $h_+$ (top) and $h_\times$  (bottom) on the periphery of the
computational domain for different positions of the initial turbulent spectrum
peak frequency $k_f / k_H = $ 300, 60, 2 from left to right respectively.
(in press)
}
\label{hhT_hhX}\end{figure*}

If significant magnetic fields exist after the phase transitions, they can
source turbulence for long durations, extending even until recombination.
For these sources, the damping due to the expansion of the universe cannot be neglected.
Numerical simulations show only a slow decay of turbulent energy, especially at the
large-scale end of the spectrum, along with the generation of significant energy
density in velocity fields; see Fig.~4 of \cite{Kahniashvili:2010gp},
and \cite{Brandenburg:2016odr}.
Turbulence in the early universe can also be generated during inflation,
whereby the magnetic field energy is injected into primordial plasma
ensuring a strong coupling between
the magnetic field and fluid motions.
The correlation scale of induced turbulent motions is limited by the
Hubble scale, as required by causality;
see \cite{Kahniashvili:2012vt} for the non-helical case
and \cite{Kahniashvili:2016bkp} for the helical case,
while the magnetic field stays frozen-in at superhorizon scales.
The strength of the turbulent motions is determined by the total energy density
of the magnetic field; a sufficiently strong field can lead to a detectable
gravitational wave signal \citep{Kahniashvili:2008pf}.

The \textsc{Pencil Code} \citep{Brandenburg:2001yb}
is a general public domain tool box to solve sets of partial differential equations on
large, massively parallel platforms.
It has recently been applied to early universe simulations of mesh size up to
$2304^3$ \citep{Brandenburg:2016odr},
which was necessary for modeling turbulence at the phase transitions
\citep{Brandenburg:2017neh} and the inflationary
stage \citep{Kahniashvili:2016bkp}.
We have recently added a module to evolve the
gravitational waves in the simulation domain
from the dynamically evolving MHD stresses.
Details of the numerical simulations can
be found in \cite{Pol:2018pao}.
Our first preliminary results are presented in Fig.~2, where we plot the gravitational
wave strain components $h_+$ and $h_\times$ sourced by fully helical hydromagnetic turbulence.
It must be highlighted that the presence of initial magnetic helicity significantly
affects the detection prospects.
However, the detection of the circular polarization degree by LISA
seems to be problematic \citep{Smith:2016jqs}.

\acknowledgments
It is our pleasure to thank the organizers, in particular Luigina Feretti and
Federica Govoni of IAU XXX FM8 ``New Insights in Extragalactic Magnetic Fields".
Partial support through the NSF Astrophysics and Astronomy Grant Program (AAG)
(1615940 \& 1615100) is gratefully acknowledged.

{}

\end{document}